\documentclass[aps,prl,showpacs,twocolumn,superscriptaddress,longbibliography]{revtex4-1}
\usepackage{graphicx,graphics,color,epsfig}
\usepackage{amsmath,amssymb,amsfonts}
\usepackage{bm,times,xspace}
\usepackage{mathrsfs,latexsym}
\usepackage{siunitx}
\usepackage[colorlinks,
citecolor=blue,
linkcolor=red,
urlcolor=blue]{hyperref}

\usepackage{dcolumn}
\usepackage{float}
\usepackage{color}
\usepackage{ulem}

\def\bs#1{\boldsymbol{#1}}

\UseRawInputEncoding

\begin{document}

\title{
Phonon Spin Selective One-Way Axial Phonon Transport in Chiral Nanohelix
}

\author{Jia Li}\thanks{J. Li and Y.-T. Tan contributed equally to this work.}
\affiliation{Key Laboratory of Jiangxi Province for Special Optoelectronic Artificial Crystal Materials, School of Chemistry and Chemical Engineering. Jinggangshan University, Ji'an, Jiangxi, 343009, P. R. China}
\affiliation{Center for Phononics and Thermal Energy Science, China-EU Joint Lab on Nanophononics, School of Physics Science and Engineering, Tongji University, 200092 Shanghai, China}

\author{Yu-Tao Tan}\thanks{J. Li and Y.-T. Tan contributed equally to this work.}
\author{Yizhou Liu}\email{Corresponding author. E-mail: yizhouliu@tongji.edu.cn}
\author{Jie Ren}\email{Corresponding author. E-mail: Xonics@tongji.edu.cn}
\affiliation{Center for Phononics and Thermal Energy Science, China-EU Joint Lab on Nanophononics, School of Physics Science and Engineering, Tongji University, 200092 Shanghai, China}

\begin{abstract}
Selectively exciting and manipulating phonons at nanoscale becomes more and more important {but still remains challenging} in modern nano-energy control and information sensing. Here, we show that the phonon spin angular momentum provides an extra degree of freedom to achieve versatile manipulation of {axial} phonons in nanomaterials via coupling to spinful multi-physical fields, such as circularly polarized infrared absorption. In particular, we demonstrate the nanoscale one-way {axial} phonon excitation and routing in chiral nanomaterials, by converting the photon spin in circularly polarized optical fields into the collective interference phonon spin. As exemplified in the smallest chiral carbon nanotube, we show that the rectification rate can reach nearly 100\%, achieving an ideal one-way phonon router, which is verified by molecular dynamics simulations. Our results shed new light on the flexible phonon manipulation via phonon spin degree of freedom, paving the way for future spin phononics.
\end{abstract}

\maketitle



\begin{figure}
    \centering
    \includegraphics[width=\linewidth]{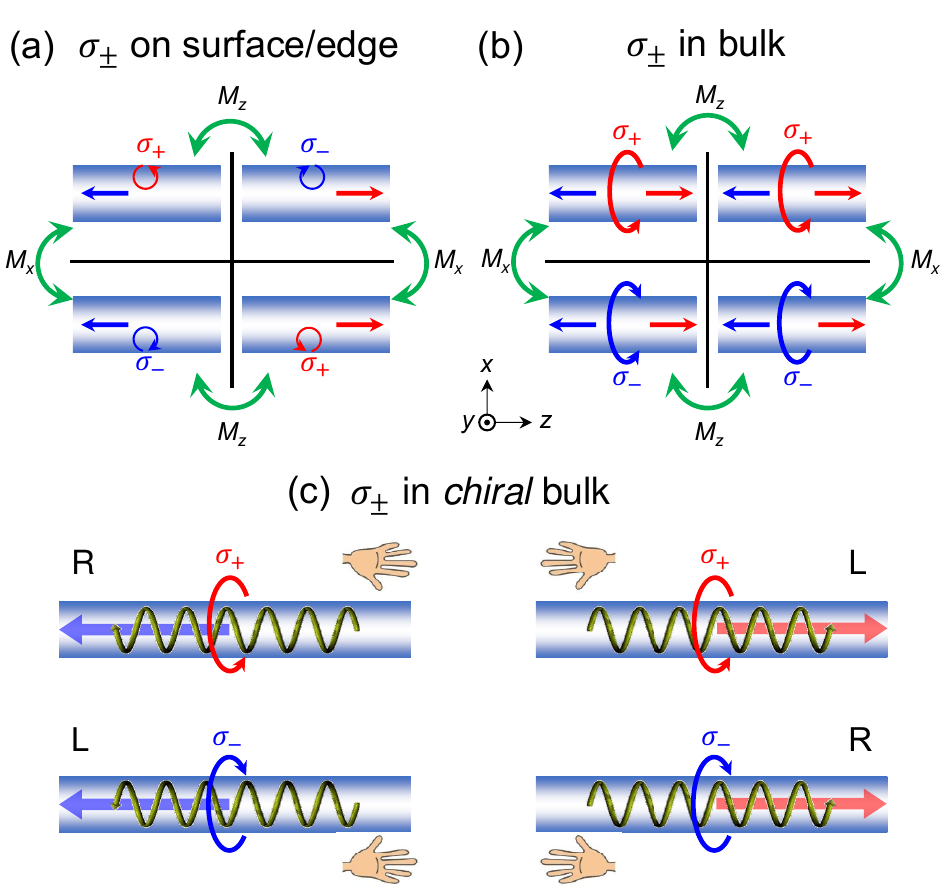}
    \caption{\textit{Symmetry transformation properties of unidirectional wave excitations.} (a) Schematics of one-way wave excited by transverse circularly polarized sources ($\sigma_\pm$) on the surface or edge. (b) For longitudinal sources $\sigma_\pm$ through the bulk, the unidirectionality is forbidden by symmetry. (c) Unidirectional wave can be excited in chiral bulk materials with longitudinal sources $\sigma_\pm$. R and L are short for right and left handedness of materials, respectively.}
    \label{fig1}
\end{figure}

\textit{Introduction.} The ability to control nanoscale flow of energy and information, particularly to realize one-way transport \cite{Li2012Jul, Maldovan2013Nov, Tokura2018Sep, Nassar2020Sep, Nagaosa2024Mar, Chang2006Nov, Yang2008Dec, Liang2010Dec, Fleury2014Jan, Xu2019Apr, Chen2022Feb, Jiang2018Dec, Xu2021May, Yang2015Mar, Liu2017Aug, Liu2019Sep}, is crucial for both fundamental science and broad practical applications. 
One typical example is the surface plasmon polariton (SPP) for manipulating optical energy at the nanoscale \cite{Ritchie1957Jun, Maier2003Apr, Willets2007May, Mayer2011Jun}. By adjusting the polarization state of the incident light \cite{Knight2007Jul, Li2010May}, the SPPs in the nanowires can be selectively excited efficiently \cite{Wei2011Feb, Zhang2011Aug, Ditlbacher2005Dec}, thereby realizing photonic logic gates \cite{Wei2011Feb, Wei2011Jul} and optical routers \cite{Wei2011Feb, Fang2010May} at nanoscale [Fig. \ref{fig1}(a)].
In crystalline solids, the lattice energies are mainly carried by vibrational waves, or say, phonons, which are difficult to control at the nanoscale due to the lack of rich tunable phonon degrees of freedom. As a result, the rectification efficiency of phononic thermal transport at nanomaterials initially only reached few percentage points \cite{Chang2006Nov}, unless based on strong heterogeneity and nonlinearity. 
Therefore, it is highly desirable to develop new means of controlling phonon flow based on new degrees of freedom. 
The recent revival of intrinsic phonon spin \cite{1961PS, 1962PS, 1962PSLevine}, from the elastic spin \cite{ElasticSpin2018PNAS, ElasticSpin2021NC, PS2022CPL, ElasticSpin2023PRL}, and acoustic spin \cite{AcousticSpin2019NSR, AcousticSpin2020NSR, AcousticSpin2020NC,WaveSpin2025CPL}, as well as the ``pseudo angular momentum'' in chiral \cite{Zhang2014Feb, Zhang2015Sep, Zhu2018Feb} {and axial phonons \cite{Juraschek2025Sep}} in crystalline solids sheds new light on this issue. One-way phonon modes could be excited through coupling with other spinful external fields and utilizing spin-momentum locking effects in various phononic systems including topological phononic materials \cite{Wang2015Sep, He2016Dec, Liu2017Dec, Liu2019Jul, ElasticSpin2022PRL, Huang2024Sep}, surface elastic waves \cite{ElasticSpin2021NC, ElasticSpin2023PRL}, topological phononic cavity-circuits \cite{ElasticSpinWG2024AS} and chiral elasticity with phononic spin-orbit interaction \cite{Yang2024Nov}.

From symmetry perspective, achieving one-way phonon transport requires breaking both time-reversal ($\mathcal{T}$) and spatial-inversion ($\mathcal{P}$) symmetries. For example, applying a transverse circularly polarized field ($\sigma_{\pm}$) onto the surface or edge [Fig. \ref{fig1}(a)] can generate transverse spin selected unidirectional elastic waves both inside the bulk \cite{Zhang2024Sep} and on the surface/interface \cite{Yuan2021Nov, Yang2023Sep, Huang2024Sep}. Nevertheless, unlike transverse $\sigma_{\pm}$, longitudinal spin cases [Fig. \ref{fig1}(b)] cannot break mirror ($\mathcal{M}_z$) or $\mathcal{P}$ symmetry along transport direction. As such, to obtain one-way excitation and routing, the requirement of $\mathcal{P}$ symmetry breaking needs to be achieved through the material itself. Chiral nanomaterials are a typical class of systems without $\mathcal{P}$ symmetry, whose band structures exhibit longitudinal-spin-momentum locking \cite{2021CISSNM, Yang2024Nov, Wan2023Feb, Ouyang2025Sep}. 
Therefore, it is crucial to identify the means of efficiently exciting one-way phonon modes at the nanoscale with exploiting phonon spin selectivity.

{In recent literature, the phonon angular momentum in crystalline solids is ad hoc defined as the sum of individual atom's angular momentum \cite{Zhang2014Feb, Juraschek2025Sep}, where the nonzero phonon angular momentum relies on the circularly vibration of individual atoms locally. However, this kind of local picture of phonon angular momentum is not consistent with the collective nature of phonon, which is defined as quasi-particles in momentum space and is nonlocal in real space \cite{PS2022CPL}. Recently, it is explicitly pointed out that there exists a nonlocal part of collective interference phonon angular momentum which can be manifested in infrared circular dichroism absorption with phonon-photon interaction  \cite{Liu2025Jun}. Therefore, many important scientific questions arise: What is the microscopic mechanism of phonon coupling to other spinful physical fields, such as circularly polarized light fields ($\sigma_\pm$), with photon-phonon spin transfer? When coupling to external spinful fields, what is the manifestation of phonon spin angular momentum? - the individual atomic angular momentum \cite{Zhang2014Feb}, or the collective interference phonon spin (CIPS) \cite{Liu2025Jun}?}

Here, we propose the effect of one-way axial phonon excitation and routing in chiral nanohelixes by circularly polarized light fields, as shown in Fig. \ref{fig1}(c). The angular momentum of the excited axial phonon is the CIPS which contains two parts: the local part and nonlocal part. The local part is the summation of individual atom' angular momentum, and the nonlocal part is the interference-induced collective phonon spin. We compare the quantitive difference between the local summation of atoms' angular momenta and CIPS in the smallest chiral carbon nanotube (CNT), and show that it is the CIPS that determines the selective excitation and transport of phonon under circularly polarized light fields. Moreover, by choosing the appropriate excitation frequency where the sign of CIPS is locked with group velocity, the one-way rectification rate in (4,2)-CNT can reach nearly 100\% - an ideal one-way phonon router. The Fourier analysis of phonon spectrum unambiguously shows that the excited unidirectional phonon modes originate from CIPS. Our finding shed new light on the flexible phonon manipulation via phonon spin dynamics, paving the way for future spin phononics.

\begin{figure}
\centering
\includegraphics[width=\linewidth]{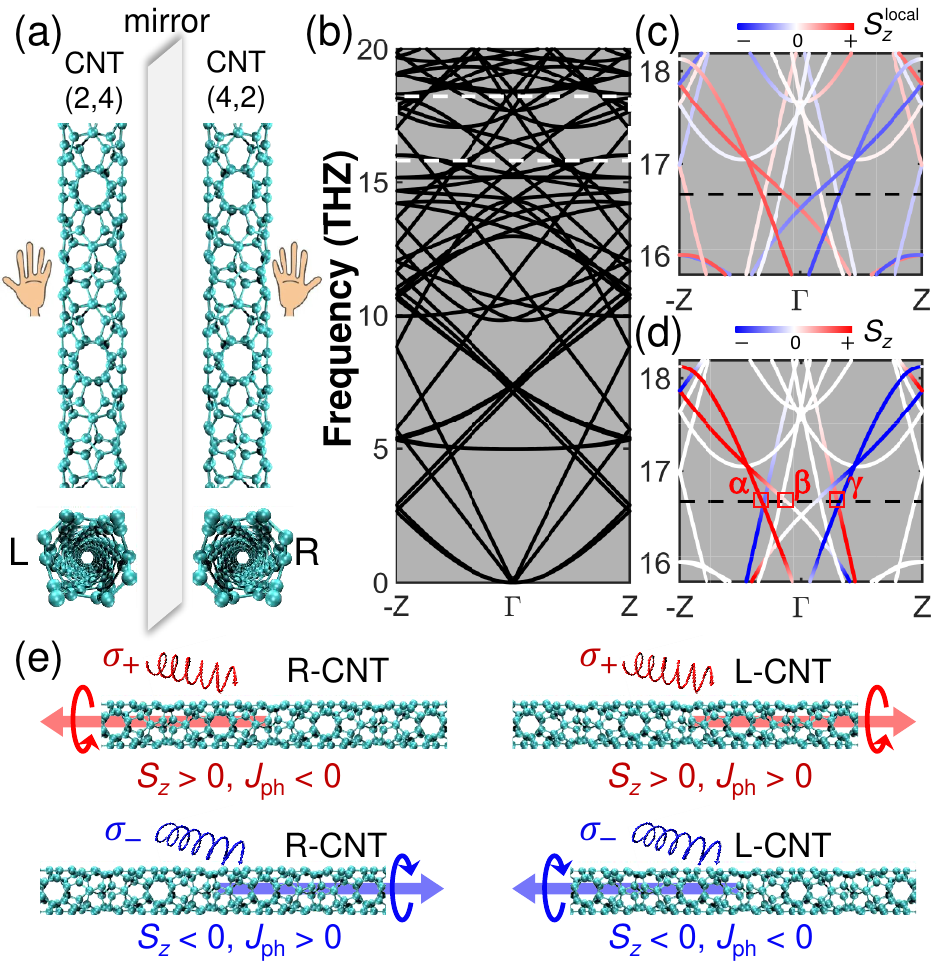}
\caption{ 
\textit{Collective interference phonon spin (CIPS) in chiral carbon nanotube (CNT).} (a) Atomic structure of CNT (4,2) (left-handed, LH) and (2,4) (right-handed, RH), respectively. (b) Calculated phonon band structure of (4,2)- and (2,4)-CNTs (see details of calculation methods in section S5 of \cite{SI}). (c) Zoom-in band structure of (b) (indicated by the white dashed box) and band-resolved $S^{\textrm{local}}_z$. (d) Same as (c) but with $S^{\textrm{local}}_z$ replaced by the CIPS $S_z$. (e) Schematics of one-way phonon routing in left- and right-haned CNT under circularly polarized light fields $\sigma_\pm$.
}
\label{fig2}
\end{figure}

\textit{CIPS in chiral CNT.} The geometry of CNTs is characterized by the chiral vector $\bs{C}_h = n \bs{a}_1 + m \bs{a}_2$ ($\bs{a}_1$ and $\bs{a}_2$ are lattice vectors of graphene) or just $(n,m)$ for short with $n$ and $m$ being two integers \cite{Saito_Dresselhaus_CNT_BOOK, Laird2015Jul}. For simplicity and without loss of generality, we consider the smallest chiral CNT, i.e. (4,2)- and (2,4)-CNTs whose geometrical structures are of opposite chirality as shown in Fig. \ref{fig2}(a). Right- (R-) and left- (L-) handed CNTs share the same phonon band [shown in Fig. \ref{fig2}(b)] but with different eigenvectors which are complex conjugate of each other as a result of $\mathcal{PT}$ symmetry. The CIPS, which describes the angular momentum of the center of mass of a unit cell, is defined as $M_\textrm{C} \bs{u} \times \dot{\bs{u}}_\textrm{C}$, where $M_\textrm{C} = \sum_i m_i$ refers to the total atomic mass of a unit cell and $\bs{u}_\textrm{C} = \sum_i \bar{m}_i \bs{u}_i$ is the center of mass displacement, with $\bar{m}_i = m_i/M_\textrm{C}$ and $\bs{u}_i$ being the reduced atomic mass and atomic displacement, respectively \cite{Liu2025Jun}. For a single-phonon mode with angular frequency $\omega$, the time-averaged value of CIPS can be calculated as 
\cite{Liu2025Jun}
\begin{equation}\label{CIPS}
    \bs{S} = \sum_{i,j} M_{ij} \langle \bs{u}_i | \hat{\bs{s}} | \bs{u}_j \rangle = \bs{S}^{\textrm{local}} + \bs{S}^{\textrm{nonlocal}}
\end{equation}
where $M_{ij} = \bar{m}_i \bar{m}_j$, $|\bs{u}_i\rangle = (u^x_i, u^y_i, u^z_i)^T$ ($T$ stands for matrix transposition) is the atomic displacement of $i$-th atom, and $\hat{\bs{s}}$ is the spin-1 operator:
\begin{equation*}
    \hat{s}_x =
    \begin{pmatrix}
    0 & 0 & 0\\
    0 & 0 & -i\\
    0 & i & 0\\
    \end{pmatrix},
    \hat{s}_y =
    \begin{pmatrix}
    0  & 0 & i\\
    0  & 0 & 0\\
    -i & 0 & 0\\
    \end{pmatrix},
    \hat{s}_z =
    \begin{pmatrix}
    0 & -i & 0\\
    i &  0 & 0\\
    0 &  0 & 0\\
    \end{pmatrix}.
\end{equation*}
The CIPS introduced in Eq. \eqref{CIPS} contains both local and nonlocal terms. The local term $\bs{S}^{\textrm{local}} = \sum_i M_{ii} \langle \bs{u}_i | \hat{\bs{s}} | \bs{u}_i \rangle$ describes the summation of each atomic rotation, while the nonlocal term $\bs{S}^{\textrm{nonlocal}} = \sum_{i\ne j} M_{ij} \langle \bs{u}_i | \hat{\bs{s}} | \bs{u}_j \rangle$ manifests the interference effect between the vibrations of the $i$- and $j$-th atoms. Previous studies usually neglect the nonlocal part. However, later we will show that the nonlocal part cannot be neglected in phonon-photon interaction processes.

Figures \ref{fig2}(c)-(d) show the band-resolved $S^{\textrm{local}}_z$ and $S_z$ of R-CNT, respectively. The $x$ and $y$ components vanish because of the skew rotation symmetry along the $z$ axis. For L-CNT, the inversion/mirror image of the R-CNT, the phonon spins $S^{\textrm{local}}_z$ and $S_z$ has a sign difference compared to R-CNT. We can see that the CIPS and its local part differs quantitatively. For example, at the frequency of $16.67$ THz (indicated by the black dashed lines) there are ten phonon modes in total with five left-moving and the other five right-moving modes. Among the five left-moving modes, their CIPS $\bs{S}_z$ or its local part $S^{\textrm{local}}_z$ mainly distribute on three modes which are labeled by $\alpha$, $\beta$, and $\gamma$, respectively. Both $S_z$ and $S^{\textrm{local}}_z$ are significant in $\alpha$, while only $S^{\textrm{local}}_z$ ($S_z$) is significant in $\beta$ ($\gamma$).

\textit{Infrared phonon excitations driven by circularly polarized light fields.} Phonon spin can couple with spin angular momentum of other physical fields via angular momentum transfer processes. We consider the process of phonon excitation by external light field through infrared interaction. Classically, each atom experiences an electrical force $\bs{F}_i = Q \bs{E}$ with $Q_i$ and $\bs{E} = (E_x, E_y, E_z)$ being the (effective) charge of $i$-th atom and the external electrical field, respectively. We consider the electrical field of the form $(E_x, E_y, E_z)=(E_0 \cos(2\pi f_0 t), E_0\cos(2\pi f_0 t+\phi_0), 0)$ applied in the central unit cell of R-CNT with frequency $f_0=16.67$ THz. Different values of $\phi_0=+\pi/2, -\pi/2, 0$ correspond to left-handed ($\sigma_-$), right-handed ($\sigma_+$), linear-polarized states of light which can couple with the chiral CNTs thus generating one-way phonons as schematically shonw in Fig. \ref{fig2}(e). 

\begin{figure}
\centering
\includegraphics[width=\linewidth]{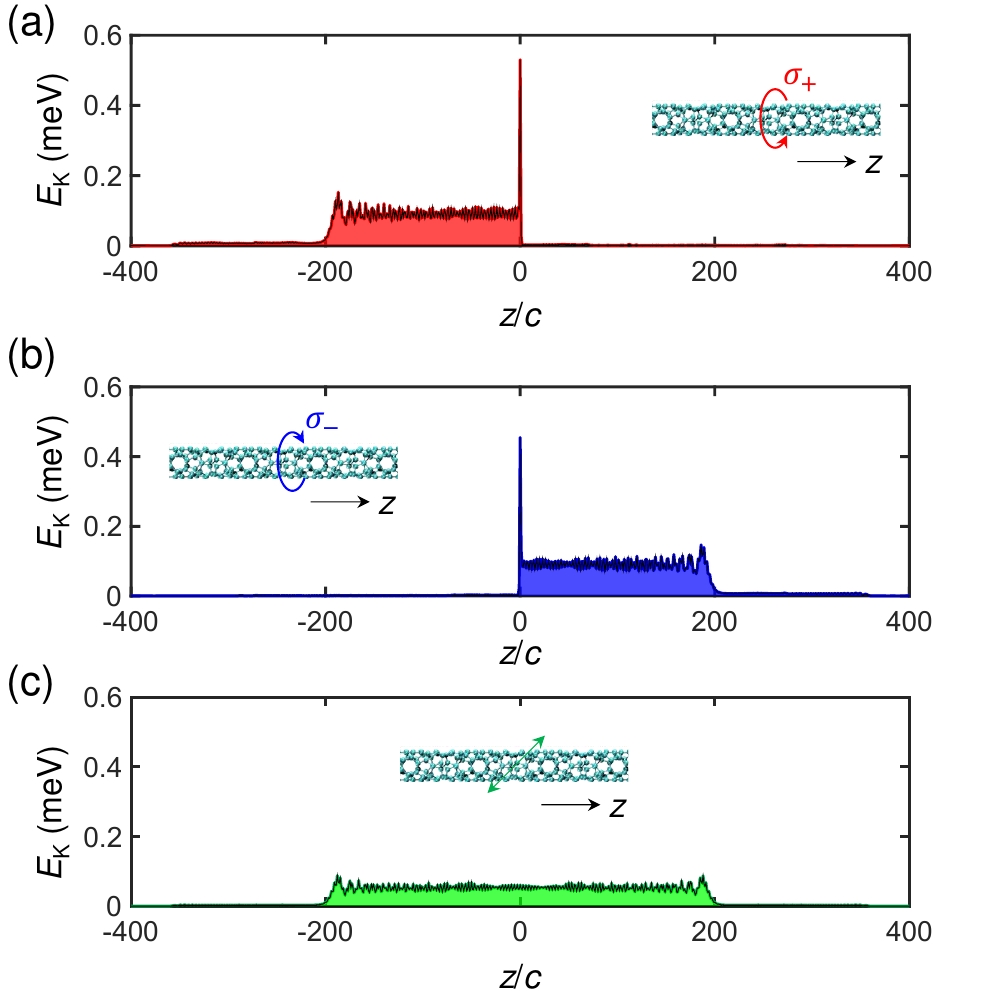}
\caption{
\textit{One-way phonon routing in (4,2)-CNT under different polarization states of light fields.} (a)-(c) Snapshot kinetic energy ($E_K$) distributions at $t=30$ ps as function of $z$ under right-handed, left-handed, and linear polarized light fields, respectively. The light fields are applied at the unit cell located at $z=0$ with $Q|E_0|=$ 1 eV/$\AA$ and $f=16.67$ THz.
}
\label{fig3}
\end{figure}

Figure \ref{fig3} shows the spatial distribution of kinetic energy under different kinds of light fields based on molecular dynamics simulations. It is clearly shown that $\sigma_+$ ($\sigma_-$) field mainly excites left- (right-) moving modes, thus realizing the one-way phonon router; while the linear-polarized field excites equal magnitudes of modes moving along opposite directions. In order to characterize the efficiency of the one-way phonon router, we investigate the diode coefficient for the energy current flow defined as
\begin{equation}
    \eta_{\pm} = \frac{J^{\sigma_\pm}_\leftarrow - J^{\sigma_\pm}_\rightarrow}{J^{\sigma_\pm}_\leftarrow + J^{\sigma_\pm}_\rightarrow} \times 100\%,
\end{equation}
where $J^{\sigma_\pm}_\leftarrow$ ($J^{\sigma_\pm}_\rightarrow$) represents the left- (right-) moving energy current under $\sigma_\pm$. The numerically calculated $\eta_+$ of (4,2) CNT can reach 95\% (see detailed theoretical analysis in sections S1-S4 of \cite{SI}) thus rationaling one-way phonon router at $f=16.67$ THz.

\begin{figure}
	\centering
	{ 
	\includegraphics[width=\linewidth]{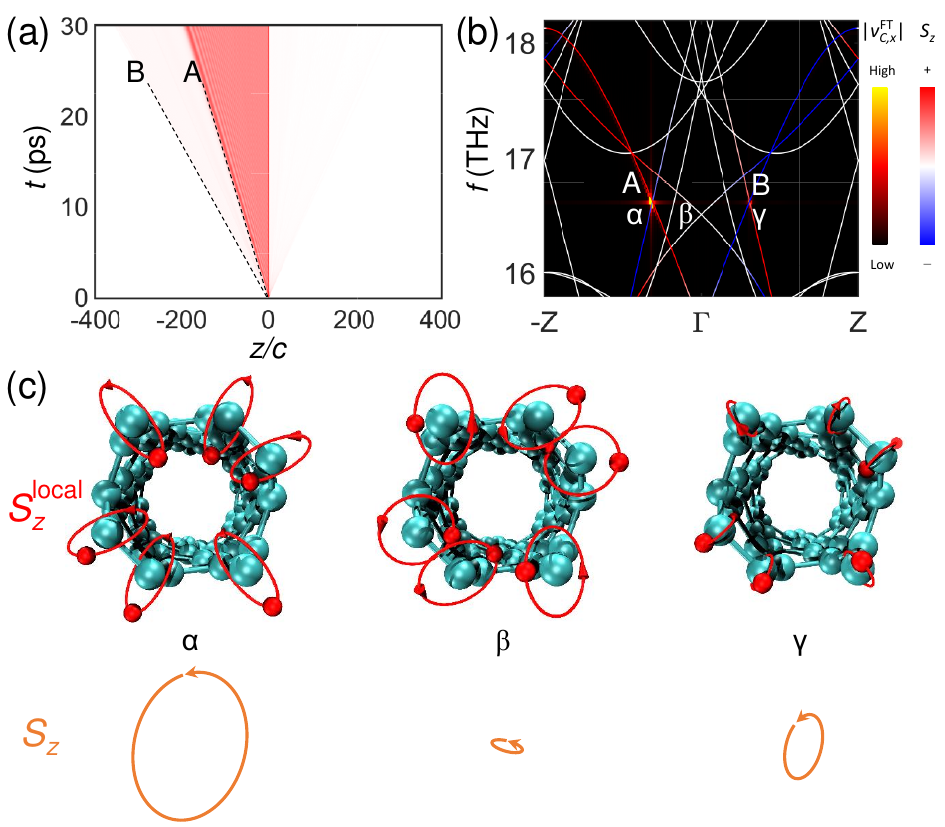}
	\caption{ 
        \textit{Space-time and momentum-frequency space distributions of excited phonon wave.} (a) Space-time distribution of $E_K$. Two modes, A (major) and B (minor), are excited with estimated group velocities 7.7 km/s and 13.6 km/s, respectively. (b) Fourier transformation (FT) of velocity of the cell-averaged center of mass $v_{C,x}$ (see details of calculation methods in section S5 of \cite{SI}) and CIPS-resolved band structure. Modes A and B coincide with $\alpha$ and $\gamma$ modes shown in Fig. \ref{fig2}(d). (c) $S^{\textrm{local}}_z$ and $S_z$ of $\alpha$, $\beta$, and $\gamma$ modes, respectively.
        }
	\label{fig4}
	}
\end{figure}

\textit{Momentum-space distribution of the excited phonon modes.} In order to understand the composition of the excited phonons, we calculate its Fourier components. Figure \ref{fig4}(a) shows the space-time distribution of the excited phonons. One can see that two modes which are labelled by A and B, respectively, are excited. The A mode (major component) has a group velocity $7.7$ km/s, and B mode (minor component) has a group velocity $13.6$ km/s. By doing Fourier transformation of the velocity of the cell-averaged center of mass, we find that there are mainly two hotspots in the $(q, f)$-space which coinsides with $\alpha$ and $\gamma$ [shown in Fig. \ref{fig4}(b)]. The cell-averaged center of mass velocity is used for Fourier analysis in order to avoid frequency-doubling problem in the kinetic energy (see details in theoretical analysis in section S4 and numerical result in section S6 of \cite{SI}). The group velocities of $\alpha$ and $\gamma$ calculated via band slope are consistent with the molecular dynamics result obtained in Fig. \ref{fig4}(a). Figure \ref{fig4}(c) shows $S^{\textrm{local}}_z$ (upper panel) and $S_z$ (lower panel) of $\alpha$, $\beta$, and $\gamma$ modes, respectively. Clearly, both $\alpha$ and $\beta$ modes have relatively larger $S^{\textrm{local}}_z$, but the CIPS $S_z$ almost vanishes in $\beta$ mode; whereas $\gamma$ mode has larger $S_z$ than $\beta$ mode although its local part is smaller. This indicates that the CIPS rather than local phonon spin plays a major role in the phonon excitation process.

\textit{Discussions.} To understand the role of CIPS in the phonon excitation process under circularly polarized light fields, we consider the the excitation rate
\begin{eqnarray}
\label{P_pm}
    P_\pm = \frac{2\pi}{\hbar} |\langle \Phi_e| H_I | \Phi_0 \rangle|^2 \delta 
    = \frac{Q^2}{\hbar^2} | \langle nq|  \bs{E}_\pm \cdot \sum_i \bs{u}_i | 0 \rangle  |^2 \tilde{\delta},
\end{eqnarray}
where $|\Phi_0\rangle = |0\rangle$ is the vacuum state without phonon excitations and $|\Phi_e\rangle = |nq\rangle$ is the one-phonon excited states of $n$-th band with wave vector $q$. $H_I = Q\bs{E}_\pm \cdot \sum_i \bs{u}_i$ is the infrared interaction Hamiltonian with $\bs{E}_\pm = \frac{E_0}{\sqrt{2}} (1, \pm i, 0)$ being the circularly polarized electrical field. $-$ ($+$) corresponds to the left (right) handedness. $\bs{u}_i$ is the atomic displacement in the middle of the CNT and the summation is over $0$-th unit cell where the electrical field is applied during the molecular dynamics simulation. $\delta = \delta(2\pi\hbar f_{nq} - 2\pi\hbar f_0)$ and $\tilde{\delta} = \delta(f_{nq} - f_0)$ refer to the law of energy conservation in the phonon-photon interaction with $f_0$ and $f_{nq}$ being the frequency of light field and phonon mode, respectively. By substituting the second quantization expression of $\bs{u}_i$ into Eq. \eqref{P_pm}, we can derive that \cite{SI}
\begin{equation}\label{dP_pm}
    P_+ - P_- = \frac{Q^2 |E_0|^2 \tilde{\delta}}{4\pi\hbar N f_0} \sum_{i,j} \langle \bs{u}^i_{nq} | \hat{s}_z | \bs{u}^j_{nq} \rangle = \frac{Q^2 |E_0|^2 \tilde{\delta}}{4\pi \hbar N f_0} \mathcal{S}_z,
\end{equation}
with $m$ being the atomic mass of carbon and $N$ being the total number of unit cells. $\mathcal{S}_z = \sum_{i,j} \langle \bs{u}^i_{nq} | \hat{s}_z | \bs{u}^j_{nq} \rangle = \mathcal{S}^{\textrm{local}}_z + \mathcal{S}^{\textrm{nonlocal}}_z$ is proportional to CIPS which contains not only local terms $\mathcal{S}^{\textrm{local}}_z = \sum_i \langle \bs{u}^i_{nq} | \hat{s}_z | \bs{u}^i_{nq} \rangle$ but also nonlocal terms $\mathcal{S}^{\textrm{nonlocal}}_z = \sum_{i\ne j} \langle \bs{u}^i_{nq} | \hat{s}_z | \bs{u}^j_{nq} \rangle$. It should be noted that the condition of momentum conservation is relaxed here because the infrared interaction is acted only at the center of the CNT. A physical quantity localized in real space has a broadband spectrum in momentum space. Therefore, the phonon modes away from $\Gamma$ can be excited. If the infrared interaction applies uniformly to all unit cells, the momentum conservation is restored \cite{Liu2025Jun}.

Equation \eqref{dP_pm} indicates that the spin angular momentum of light is transferred to CIPS but not the simple summation of individual atomic angular momentum. The nonlocal terms in CIPS reflects the wave nature of phonon. According to the principles of particle-wave duality and quantum uncertainty, each (quasi) particle with fixed energy and momentum corresponds to an extended wave in space-time coordinates. Therefore, in the real space many atoms couple with external fields simultaneously and the collective nature plays crucial roles there. In the phonon-photon interaction processes, the vibrations of all atoms couples with the electrical field simultaneously, and it is the collective motion of the lattice vibration rather than the individual atomic motion that interacts with the light fields.

\textit{Summary and outlook.} In this work, as exemplified in (4,2)-CNT, we unvealed the pivotal role of collective interference phonon spin (CIPS) in the phonon-photon interaction process. The absorption rate difference between left- and right-handed circularly polarized light ($\sigma_\pm$), i.e. circular dichroism, results in angular momentum transfer from photon to phonon, which is responsible for the unidirectional phonon routing effect due to the spin-momentum locking effect in chiral materials.

Phonons play a major role in many physical processes such as Raman scattering, electron-phonon coupling, and magnon-phonon coupling etc. It would be interesting to explore the role of CIPS in other phonon-involved quantum angular momentum transfer processes \cite{Lei2025Jul}, such as circularly polarized Raman scattering, electron-phonon coupling, and phonon-magnon coupling etc.

\textit{Acknowledgment.} We acknowledge the support from the National Key R\&D Program of China (No. 2022YFA1404400 and No. 2023YFA1406900), the Natural Science Foundation of Shanghai (No. 23ZR1481200), the Program of Shanghai Academic Research Leader (No. 23XD1423800), National Natural Science Fundation of China (Grant No. 21965015), Shanghai Pujiang Program (Grant No. 23PJ1413000), National Natural Science Fundation of China (Grant No. 12404279) and the Opening Project of Shanghai Key Laboratory of Special Artificial Microstructure Materials and Technology.


\providecommand{\noopsort}[1]{}\providecommand{\singleletter}[1]{#1}%

\end{document}